\documentclass[10pt, conference, compsocconf]{IEEEtran}

\usepackage{algorithm}
\usepackage{amsfonts}
\usepackage{array}
\usepackage{dgleich-math}
\usepackage{algpseudocode}
\usepackage{amsmath}
\usepackage{graphicx}
\usepackage{url}
\usepackage{tabularx}
\usepackage{booktabs}
\usepackage{listings}
\usepackage{xcolor}
\usepackage{lmodern}
\usepackage{epstopdf}

\renewcommand{\mat}[1]{#1}

\definecolor{thegreen}{rgb}{0,.5,0}
\definecolor{theblue}{rgb}{0,0,.8}
\newcommand{\ssum}[3]{\displaystyle\sum\limits_{#1}^{#2} #3}

\usepackage[tight,footnotesize]{subfigure}

\begin{document}
\title{Direct QR factorizations for tall-and-skinny \\
matrices in MapReduce architectures}


\author{\IEEEauthorblockN{Austin R. Benson}
\IEEEauthorblockA{Institute for Computational and\\
Mathematical Engineering\\
Stanford University\\
arbenson@stanford.edu}
\and
\IEEEauthorblockN{David F. Gleich}
\IEEEauthorblockA{Department of Computer Science\\
Purdue University\\
dgleich@purdue.edu}
\and
\IEEEauthorblockN{James Demmel}
\IEEEauthorblockA{Computer Sciences Division and\\
Department of Mathematics \\
University of California, Berkeley\\
demmel@cs.berkeley.edu}
}

\maketitle

\begin{abstract}
The QR factorization and the SVD are two fundamental matrix decompositions with applications throughout scientific computing and data analysis.  For matrices with many more rows than columns, so-called ``tall-and-skinny matrices,'' there is a numerically stable, efficient, communication-avoiding algorithm for computing the QR factorization.  It has been used in traditional high performance computing and grid computing environments.  For MapReduce environments, existing methods to compute the QR decomposition use a numerically unstable approach that relies on indirectly computing the Q factor.  In the best case, these methods require only two passes over the data. 
In this paper, we describe how to compute a stable tall-and-skinny QR factorization on a MapReduce architecture in only slightly more than 2 passes over the data.  We can compute the SVD with only a small change and no difference in performance.  We present a performance comparison between our new direct TSQR method, a standard unstable implementation for MapReduce (Cholesky QR), and the classic stable algorithm implemented for MapReduce (Householder QR).  We find that our new stable method has a large performance advantage over the Householder QR method.  This holds both in a theoretical performance model as well as in an actual implementation.
\end{abstract}


%
\IEEEpeerreviewmaketitle

\section{Introduction}
The QR factorization of an $m \times n$ real-valued matrix $A$ is: 
\[ A = QR \]
where $Q$ is an $m \times n$ orthogonal matrix and $R$ is an $n \times n$ upper triangular matrix.  We call a matrix tall-and-skinny if it has many more rows than columns $(m \gg n)$.  In this paper, we study algorithms to compute a QR factorization of a tall-and-skinny matrix for nearly-terabyte sized matrices on MapReduce architectures~\cite{Dean2004-MapReduce}. Current tall-and-skinny QR methods for MapReduce provide \emph{only} a fast way to compute $R$~\cite{Constantine-2011-TSQR-MapReduce}. (The details of these are described further in Sec.~\ref{sec:indirect_QR}.)  In order to compute the matrix $Q$, they use the indirect formulation: 
\[ Q = AR^{-1}. \]
For $R$ to be invertible, $A$ must be full-rank, and we assume $A$ is full-rank throughout this paper.  The indirect formulation is known to be numerically unstable, although, a step of iterative refinement can sometimes be used to produce a $Q$ factor with acceptable accuracy~\cite{parlett1998}.  (Iterative refinement is the process of repeating the QR decomposition on the computed $Q$ factor.)  However, if a matrix is sufficiently ill-conditioned, iterative refinement will still result in a large error measured by $\normof[2]{Q^TQ - I}$ (see Sec.~\ref{sec:stability}).  We shall describe a numerically stable method (Sec.~\ref{sec:direct_QR}) that computes $Q$ and $R$ directly and faster than performing the refinement of the indirect computation for some matrices.

Sec.~\ref{sec:perf_model} describes a performance model for our algorithms, which allows us to compute lower bounds on running times.  The algorithms are almost always within a factor of two of the lower bounds (Sec.~\ref{sec:perf_comparison}).


\subsection{MapReduce motivation}

The data in a MapReduce computation is defined by a collection of key-value pairs. When we use MapReduce to analyze tall-and-skinny matrix data, a key represents the identity of a row and a value represents the elements in that row.  Thus, the matrix is a collection of key-value pairs.  We assume that each row has a distinct key for simplicity; although we note that our methods also handle cases where each key represents a set of rows.  

There are a growing number of MapReduce frameworks that implement the same computational engine: first, \emph{map} applies a function to each key-value pair which outputs a transformed key-value pair; second, \emph{shuffle} rearranges the data to ensure that all values with the same key are together; finally, \emph{reduce} applies a function to all values with the same key.  The most popular MapReduce implementation -- Hadoop~\cite{Hadoop2012-021} -- stores all data and intermediate computations on disk. Thus, we do not expect numerical linear algebra algorithms for MapReduce to be faster than state-of-the-art in-memory MPI implementations running on clusters with high-performance interconnects.  However, the MapReduce model offers several advantages that make the platform attractive for large-scale, large-data computations (see also~\cite{Zhao2009-MapReduce} for information on tradeoffs).  
First, many large datasets are already warehoused in MapReduce clusters.  With the availability of algorithms, such as QR, on a MapReduce cluster, these data do not need to be transferred to another  cluster for analysis.  Second, MapReduce systems like Hadoop provide transparent fault-tolerance, which is a major benefit over standard MPI systems.  Other MapReduce implementations, such as Twister~\cite{Ekanayake-2010-Twister}, Phoenix++~\cite{Talbot-2011-pheonix++}, LEMOMR~\cite{Fadika-2011-benchmarking}, and MRMPI~\cite{Plimpton-2011-MRMPI}, often store data in memory and may be a great deal faster; although, they usually lack the automatic fault tolerance.  Third, the Hadoop computation engine handles all details of the distributed input-output routines, which greatly simplifies the resulting programs.

For the majority of our implementations, we use Hadoop streaming and the Python-based Dumbo MapReduce interface~\cite{Dumbo2012}. These programs are concise, straightforward, and easy-to-adapt to new applications.  We have also investigated C++ and Java implementations, but these programs offered only mild speedups (around 2-fold), if any. See Table~\ref{tab:perf_cxx} for a comparison against C++.  
The Python implementation uses about 70 lines of code, while the C++ implementation uses about 600 lines of code.  

\begin{table}[tbp]
\vspace{-\baselineskip}
\centering
\caption{The performance improvement of C++ over Python for our  Direct TSQR on a 10-node MapReduce cluster is only mild.}
\begin{tabular}{l l @{\qquad} c c}
\toprule
Rows & Cols. & Job time & Speedup \\
           &            &  (secs.)   &                \\
\midrule
4,000,000,000 & 4     & 2217 & 2.76 \\
2,500,000,000 & 10   & 3137 & 1.29 \\
600,000,000    & 25   & 1482 & 1.29 \\
500,000,000    & 50   & 1477 & 2.09 \\
150,000,000    & 100 & 1503 & 1.43 \\
\bottomrule
\end{tabular}
\label{tab:perf_cxx}
\end{table}

\subsection{Success metrics}
Our two success metrics are speed and stability.  The differences in speed are examined in Sec.~\ref{sec:perf_comparison}.  To analyze the performance, we construct a performance model for the MapReduce cluster.  After fitting two parameters to the performance of the cluster, it predicts the runtime to within a factor of two.  For stability, we use the metric $\normof[2]{A - QR}/\normof[2]{R}$ to measure the accuracy of the decomposition and $\normof[2]{Q^TQ - I}$ to measure the orthogonality of the computed $Q$ factor.  Small scale simulations of the MapReduce algorithms show that, regardless of the algorithm, $\normof[2]{A - QR}/\normof[2]{R}$ is O($\epsilon$) where $\epsilon$ is the machine precision.  However, $\normof[2]{Q^TQ - I}$ varies dramatically based on the algorithm, but is always $O(\epsilon)$ for our new direct TSQR method.  We examine these differences in Sec.~\ref{sec:stability}.

\section{Indirect QR factorizations in MapReduce}\label{sec:indirect_QR}

One of the first papers to explicitly discuss the QR factorization on MapReduce architectures was written by Constantine and Gleich~\cite{Constantine-2011-TSQR-MapReduce}; however many had studied methods for \emph{linear regression} and \emph{principal components analysis} in MapReduce~\cite{Chu-2006-MapReduce}.  These methods all bear a close resemblance to the Cholesky QR algorithm we describe next.

\subsection{Cholesky QR}

The Cholesky factorization of an $n \times n$ symmetric positive definite real-valued matrix $A$ is: 
\[ A = LL^T \]
where $L$ is an $n \times n$ lower triangular matrix.  Note that, for any $\mA$ that is full rank, $A^TA$ is symmetric positive definite.  The Cholesky factor $L$ for the matrix $A^T A$ is exactly the matrix $R$ in the QR factorization as the following derivation shows.  Let $A = QR$.  Then
\[A^TA = (QR)^TQR = R^TQ^TQR = R^TR.\]
Since $R$ is upper triangular and $L$ is unique, $R^TR = LL^T$.  The method of computing $R$ via the Cholesky decomposition of $A^TA$ matrix is called \emph{Cholesky QR}.

Thus, the problem of finding $R$ becomes the problem of computing $A^T A$.  This task is straightforward in MapReduce.  In the map stage, each task collects rows -- recall that these are key-values pairs -- to form a local matrix $A_p$ and then computes $A_p^T A_p$.  These matrices are small, $n \times n$, and are output by row.  In fact, $A_p^T A_p$ is symmetric, and there are ways to reduce the computation by utilizing this symmetry. We do not exploit them because disk access time dominates the computation; a more detailed performance discussion is in Sec.~\ref{sec:perf}.
In the reduce stage, each individual reduce function takes in multiple instances of each row of $A^T A$ from the mappers.  These rows are summed to produce a row of $A^T A$.  Formally, this method computes: 
\[ A^T A = \sum_{p=1}^{P} A_p^T A_p \]
where $A_i$ is the input to each map-task.
Alg.~\ref{alg:MRAtA} explicitly shows how this is done with key-value pairs in a MapReduce architecture.

\begin{figure*}
\centering
\includegraphics[width=1.3\columnwidth]{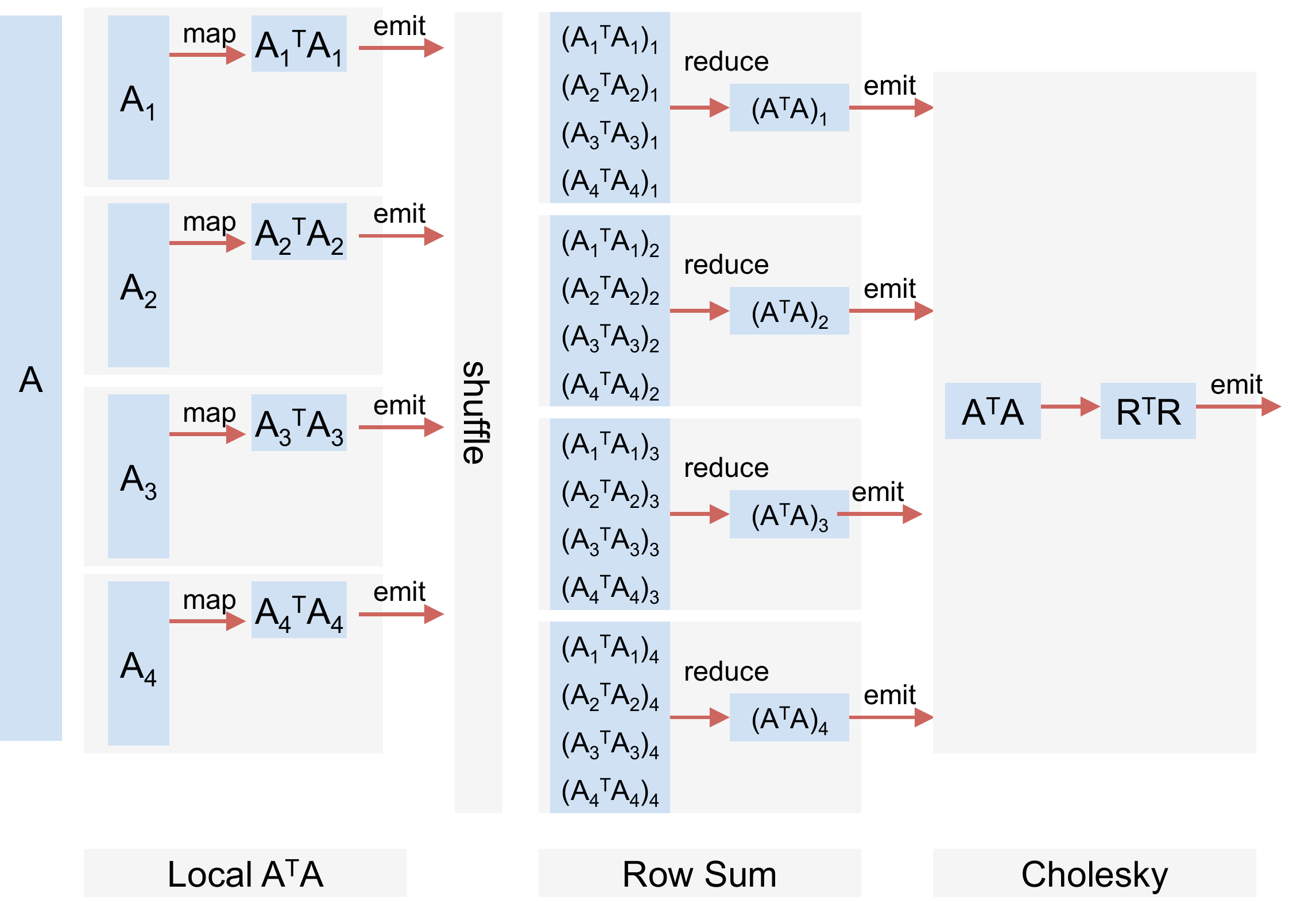}
\caption{MapReduce Cholesky QR computation for a matrix $A$ with 4 columns.}
\label{fig:MR_chol}
\end{figure*}

Extending the $A^TA$ computation to Cholesky $QR$ simply consists of gathering all rows of $A^TA$ on one processor and serially computing the Cholesky factorization $A^TA = LL^T$.  The serial Cholesky factorization is fast since $A^TA$ is small, $n \times n$.  The Cholesky $QR$ MapReduce algorithm is illustrated in Fig.~\ref{fig:MR_chol}.

\begin{algorithm}
  \caption{Compute $A^TA$ in MapReduce}
  \label{alg:MRAtA}
  \begin{algorithmic}
  \Function{map}{key $k$, val $a$} 
    \For{i, row in enumerate($a^Ta$)}
     \State emit(i, row)
    \EndFor
  \EndFunction

  \vspace{0.1in}

  \Function{reduce}{key $k$, $\langle$ vals  $v^{k}_j$ $\rangle$}
    \State emit($k$, sum($\langle v^{k}_j \rangle$))
  \EndFunction
  
  \end{algorithmic}
\end{algorithm}

It is important to note the architecture limitation due to the number of columns, $n$.  The number of keys emitted by each map task is exactly $n$: $0$, $1$, ... $n-1$ (one for each row of $A_p^TA_p$), and the total number of unique keys passed to the reduction stage is $n$.  Thus, the row sum reduction stage can use at most $n$ tasks.

Alternatively, the reduce function can emit a key-value pair where the key represents the row and column index of a given entry of $A_p^TA_p$, and the value is the given entry.  This increases the number of unique keys to $n^2$ (or, by taking symmetry into account, $n(n-1)$).  It is also valid to use more general reduction trees where partial row sums are computed on all the processors, and a reduction to $n$ processors accumulates the partial row sums.  The cost of this more general tree is the startup time for another map and reduce iteration.  Typically, the extra startup time is more expensive than the performance penalty of having less parallelism.

Each of these variations of Cholesky QR can be described by our performance model in Sec.~\ref{sec:perf_model}.  For experiments, we use a small cluster (where at most 40 reduce tasks are available), and these design choices have little effect on the running times.  We use the implementation described in Alg.~\ref{alg:MRAtA} as it is the simplest.

\subsection{Indirect TSQR}\label{sec:indir_tsqr}

One of the problems with Cholesky QR is that the matrix $A^T A$ has the \emph{square} of the condition number of the matrix $A$. This suggests that finite precision computations with $A^T A$ will not always produce an accurate $R$ matrix. For this reason, Constantine and Gleich studied a succinct MapReduce implementation~\cite{Constantine-2011-TSQR-MapReduce} of the TSQR algorithm by Demmel et al.~\cite{demmel-2008-caqr}, where map and reduce tasks both compute local QR computations.  This method is known to be numerically stable~\cite{demmel-2008-caqr} and was recently shown to have superior stability to many standard algorithms~\cite{Mori-2012-allreduce}.  Constantine and Gleich's initial implementation is only designed to compute $R$.  We will refer to this method as ``Indirect TSQR'', because $Q$ may be computed indirectly with $Q = AR^{-1}$.  In the following section, we extend this method to also compute $Q$ in a stable manner. 

\begin{figure*}
\centering
\includegraphics[width=1.2\columnwidth]{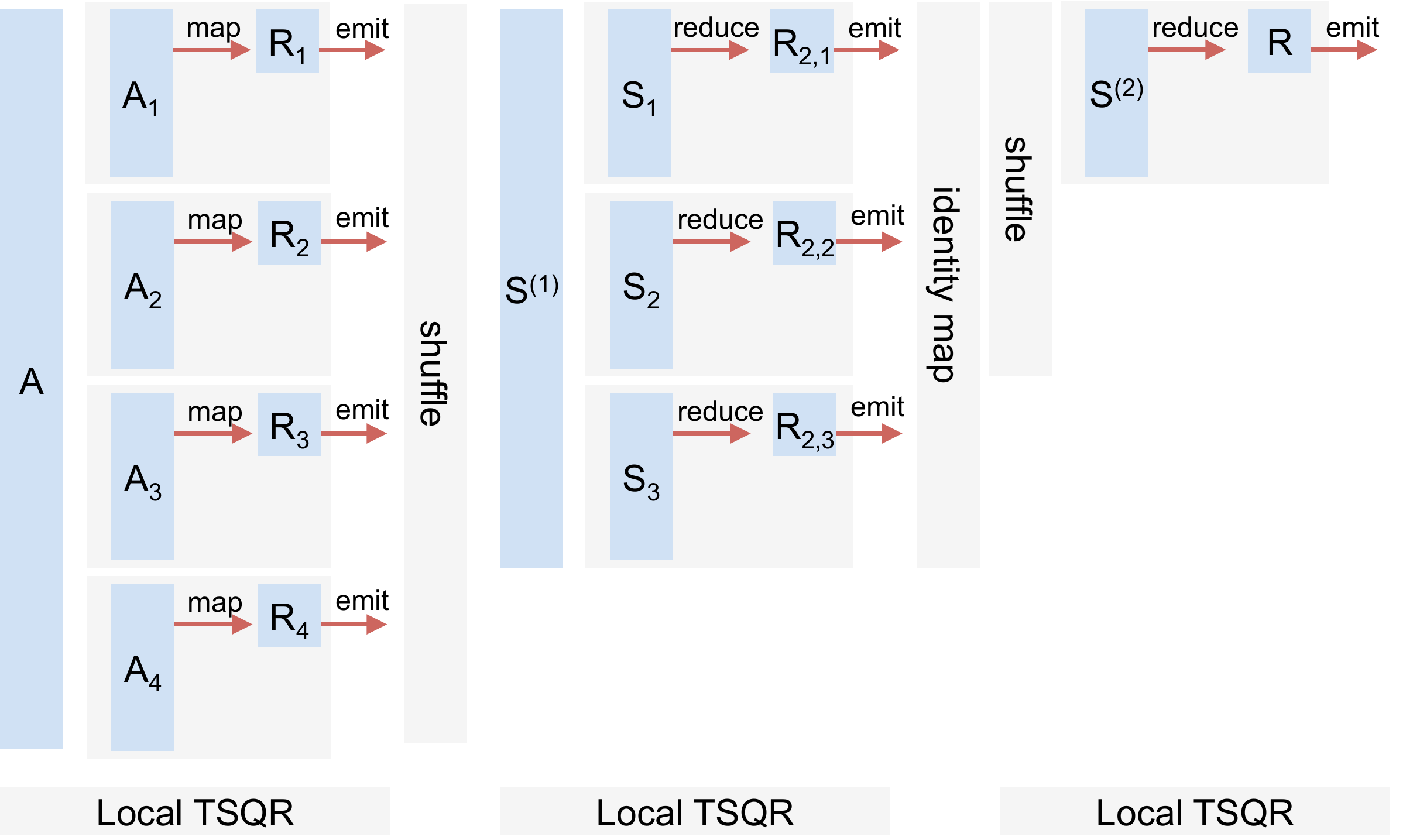}
\caption{MapReduce TSQR computation.  $S^{(1)}$ is the matrix consisting of the rows of the $R_i$ factors stacked on top of each other, $i = 1, 2, 3, 4$.  Similarly, $S^{(2)}$ is the matrix consisting of the rows of the $R_{2, j}$ factors stacked on top of each other, $j = 1, 2, 3$.}
\label{fig:MR_tsqr}
\end{figure*}

We will now briefly review the Indirect TSQR algorithm and its implementation to facilitate the explanation of the more intricate direct version.  Let $A$ be a matrix with $8n$ rows and $n$ columns, which is partitioned across four map tasks as:
\[ 
\mA = \bmat{\mA_1 \\ \mA_2 \\ \mA_3 \\ \mA_4}.
\]
Each map task computes a local $QR$ factorization:
\[ 
\mA = \underbrace{\bmat{ \mQ_1 \\ & \mQ_2 \\ & & \mQ_3 \\ & & & \mQ_4 }}_{8n \times 4n}
      \underbrace{\bmat{ \mR_1 \\ \mR_2 \\ \mR_3 \\ \mR_4}}_{4n \times n}.
\]
The matrix of stacked upper triangular matrices on the right is then passed to a reduce task and factored into $\tilde{Q}\tilde{R}$.  At this point, we have the QR factorization of $A$ in product form: 
\[ 
 \mA = \overbrace{\underbrace{\bmat{ \mQ_1 \\ & \mQ_2 \\ & & \mQ_3 \\ & & & \mQ_4 }}_{8n \times 4n}
       \underbrace{\tilde{\mQ}}_{4n \times n}}^{= \mQ} 
       \quad
       \overbrace{\vphantom{\bmat{ \mQ_1 \\ & \mQ_2 \\ & & \mQ_3 \\ & & & \mQ_4 }}\underbrace{\tilde{\mR}}_{n \times n}}^{= \mR}.
\]
The Indirect TSQR method ignores the intermediate $Q$ factors and simply outputs the $n \times n$ factors $R_i$ in the intermediate stage and $\tilde{R}$ in the final stage.  Fig.~\ref{fig:MR_tsqr} illustrates each map and reduce output.  We do not need to gather all $R$ factors onto a single task to compute $\tilde{R}$.  Any reduction tree computes $\tilde{R}$ correctly.  Constantine and Gleich found that using an additional MapReduce iteration to form a more parallel reduction tree could greatly accelerate the method. This finding differs from the Cholesky QR method, where additional iterations rarely helped. In the next section, we show how to save the $Q$ factors to reconstruct $Q$ directly.

\subsection{Computing $AR^{-1}$}
\begin{figure}[tbp]
\centering
\includegraphics[width=\columnwidth]{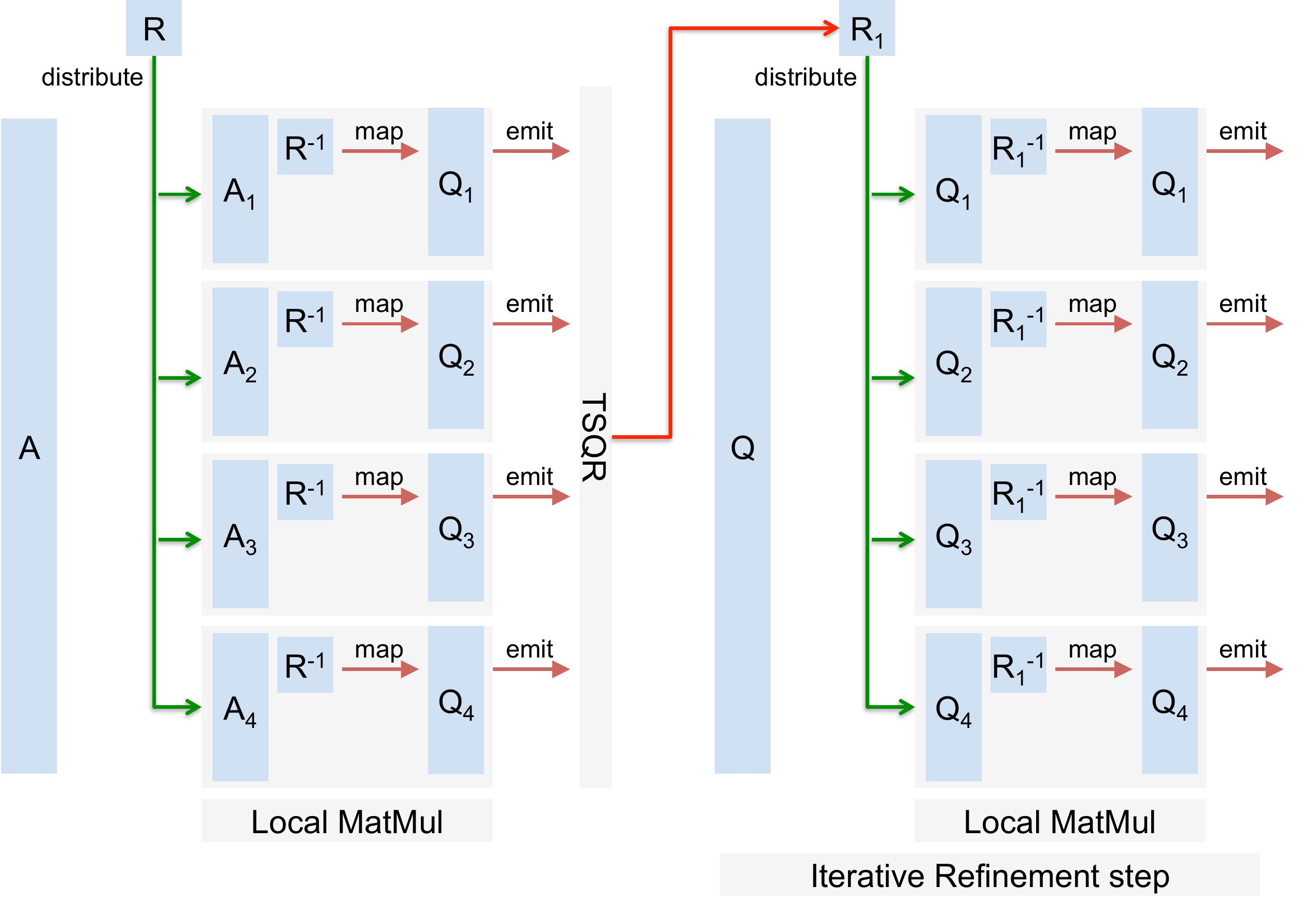}
\caption{Indirect MapReduce computation of $Q$ with iterative refinement.}
\label{fig:MR_IR}
\end{figure}

Given the matrix $R$, the simplest method for computing $Q$ is computing the inverse of $R$ and multiplying by $A$, that is, computing $AR^{-1}$.  Since $R$ is $n \times n$ and upper-triangular, we can compute its inverse quickly. Fig.~\ref{fig:MR_IR} illustrates how the matrix multiplication and iterative refinement step cleanly translate to MapReduce.  This ``indirect'' method of the inverse computation is not backwards stable (for example, see~\cite{Stathopoulos2001-QR}).  Thus, a step of iterative refinement may be used to get $Q$ within desired accuracy.  However, the indirect methods may still have large errors after iterative refinement if $A$ is ill-conditioned enough.  This further motivates the use of a direct method.

\section{Direct QR Factorizations in MapReduce}\label{sec:direct_QR}

One of the textbook algorithms to compute a stable QR factorization
is the Householder QR method~\cite{golub1996-matrix}.  This method
always produces a matrix $\mQ$ where $\normof[2]{\mQ^T \mQ - \mI}$ is on 
the order of machine error.  We begin our discussion by explaining
how to implement this method in MapReduce.

\subsection{Householder QR}

\begin{figure*}
\centering
\includegraphics[width=1.5\columnwidth]{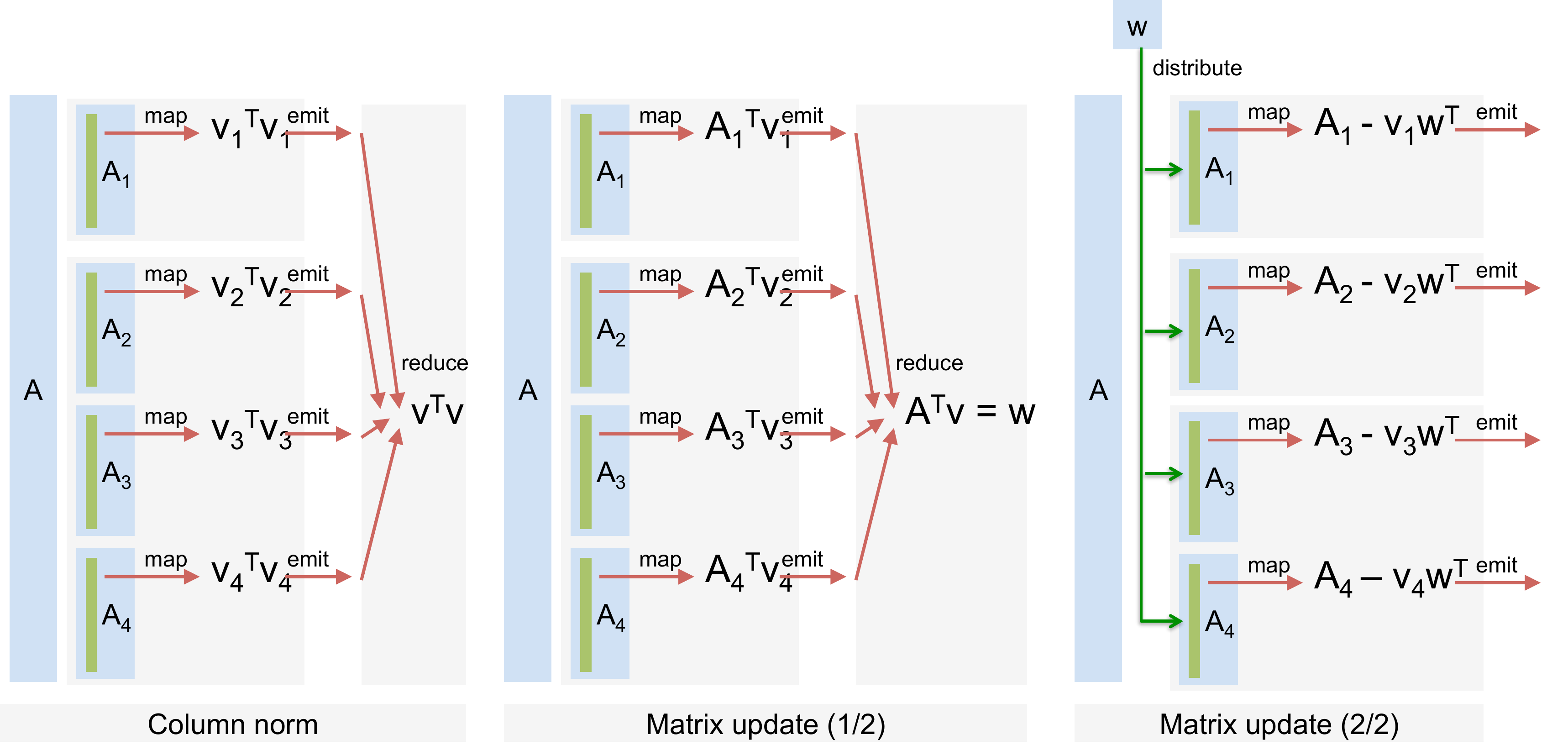}
\caption{Outline of MapReduce Householder QR.}
\label{fig:MR_householder}
\end{figure*}

The Householder QR algorithm~\cite{Trefethen1997-book} is not as friendly to MapReduce as either Cholesky QR or Indirect TSQR.  One reason for this phenomena is the iterative nature of the algorithm.  At each step of the algorithm, the matrix $A$ is completely updated.  In MapReduce, this means we must constantly rewrite the matrix on disk.  Conceptually, each step of the Householder QR method corresponds to three MapReduce calls.  These are illustrated in Fig.~\ref{fig:MR_householder}.  The first step of the algorithm computes the norm of a column of $A$ to help form the Householder reflector.  The second and third steps of the algorithm update the matrix with $A \leftarrow A - 2v(A^Tv)^T$, where $v$ is the Householder reflector.  However, in the actual implementation, the first and third steps are combined because we can compute the norm for the next step immediately after updating the matrix in the third step.  

Thus, the MapReduce Householder QR algorithm uses $2n$ passes over the data for a matrix $A$ with $n$ columns.  Every other pass requires rewriting the matrix on disk.  As $n$ grows, the performance of this algorithm becomes significantly worse than our other algorithms.

This MapReduce implementation of Householder QR is a BLAS 2 algorithm, whereas standard Sca/LAPACK uses a BLAS 3 algorithm~\cite{lapack-guide-1999,choi-1995-scalapack}.  The central reason for this is the row-wise layout of the matrix in the Hadoop Distributed File System (HDFS).  For tall-and-skinny matrices, the canonical key-value pair stored in HDFS uses a row as the matrix as the value and a unique row identifier for the key.  Thus, reading the leading columns of the matrix has the same cost as reading the entire matrix.  The stock BLAS 3 algorithm for LAPACK is a much better choice for their column-wise matrix layout.

\subsection{Direct TSQR}\label{sec:direct_tsqr}

\begin{figure*}
\centering
\includegraphics[width=1.7\columnwidth]{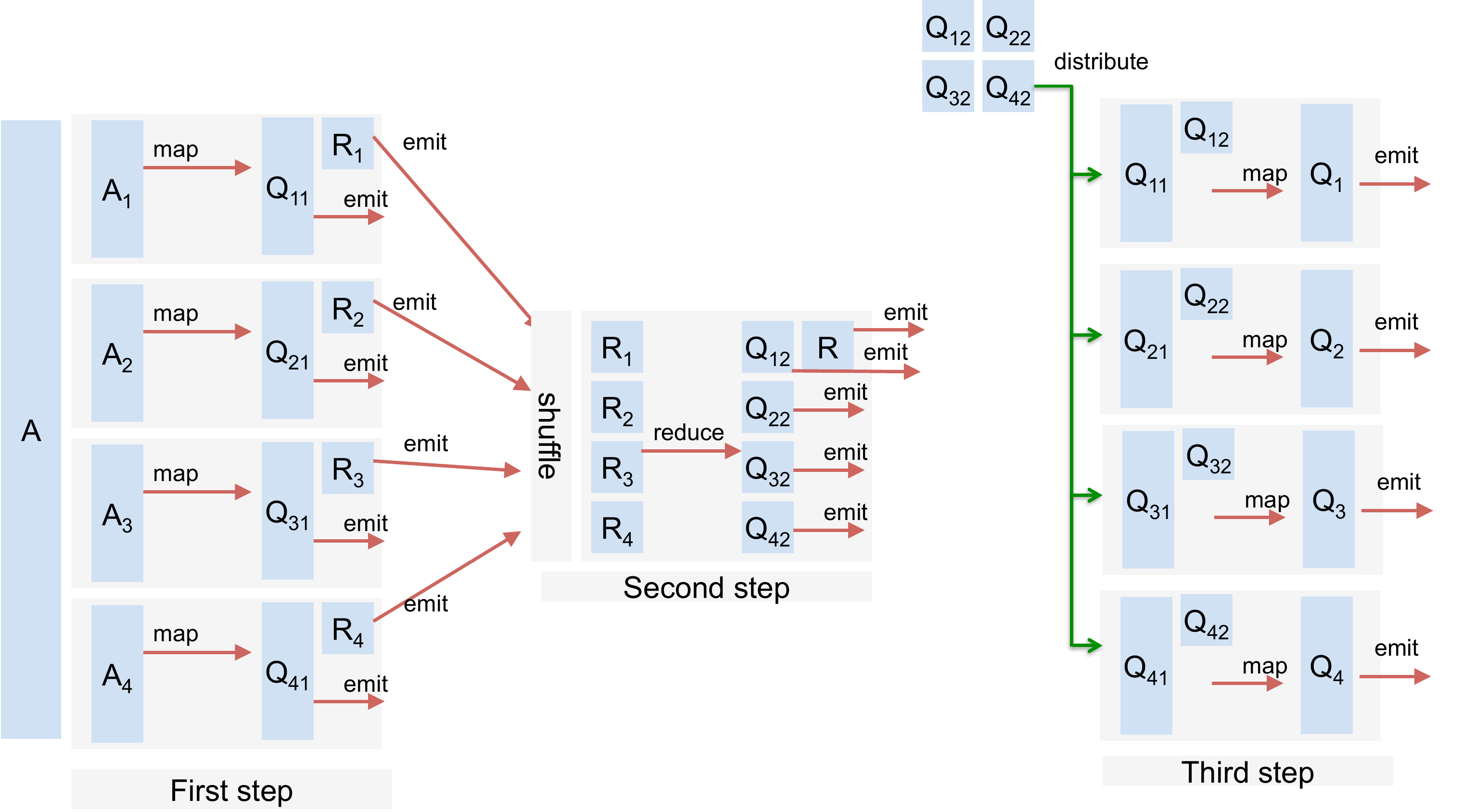}
\caption{Direct MapReduce computation of $Q$ and $R$.}
\label{fig:MR_full}
\end{figure*}

We finally arrive at our proposed method.  Here, we directly compute the QR decomposition of $A$ in three steps using two map functions and one reduce function, as illustrated in Fig.~\ref{fig:MR_full}.  This avoids the iterative nature of the Householder methods.  For an example, consider again a matrix $A$ with $8n$ rows and $n$ columns, which is partitioned across four map tasks for the first step:
\[ 
\mA = \bmat{\mA_1 \\ \mA_2 \\ \mA_3 \\ \mA_4}. 
\]
The first step uses only map tasks.  Each task collects data as a local matrix, computes a single QR decomposition, and emits $Q$ and $R$ to separate files.  The factorization of $A$ then looks as follows, with $Q_jR_j$ the computed factorization on the $j$th task:
\[ 
\mA = \underbrace{\bmat{ \mQ_1 \\ & \mQ_2 \\ & & \mQ_3 \\ & & & \mQ_4 }}_{8n \times 4n}
      \underbrace{\bmat{ \mR_1 \\ \mR_2 \\ \mR_3 \\ \mR_4}}_{4n \times n} .
\]
The second step is a single reduce task.  The input is the set of $R$ factors from the first step.  The $R$ factors are collected as a matrix and a single QR decomposition is performed.  The sections of $Q$ corresponding to each $R$ factor are emitted as values.  In the following figure, $\tilde{R}$ is the final upper triangular factor in our QR decomposition of $A$:
\[
\underbrace{\bmat{ \mR_1 \\ \mR_2 \\ \mR_3 \\ \mR_4}}_{4n \times n} = 
\underbrace{\bmat{ \mQ^2_1 \\ \mQ^2_2 \\ \mQ^2_3 \\ \mQ^2_4}}_{4n \times n}
\underbrace{\tilde{\mR}}_{n \times n} .
\]
The third step also uses only map tasks.  The input is the set of $Q$ factors from the first step.  The $Q$ factors from the second step are small enough that we distribute the data in a file to all map tasks.  The corresponding $Q$ factors are multiplied together to emit the final $Q$:
\[
 \underbrace{\mQ}_{8n \times n} = \underbrace{\bmat{ \mQ_1 \\ & \mQ_2 \\ & & \mQ_3 \\ & & & \mQ_4 }}_{8n \times 4n}
           \underbrace{\bmat{ \mQ^2_1 \\ \mQ^2_2 \\ \mQ^2_3 \\ \mQ^2_4}}_{4n \times n}
\]
\[
           = \underbrace{\bmat{ \mQ_1\mQ^2_1 \\ \mQ_2\mQ^2_2 \\ \mQ_3\mQ^2_3 \\ \mQ_4\mQ^2_4}}_{8n \times n} .
\]
\[
A = Q\tilde{R}
\]
To compute the SVD of $A$, we modify the second step and add a fourth step.  In the second step, we also compute $R = U\Sigma V^T$.  Then $A = (QU)\Sigma V^T$ is the SVD of A.  Since $R$ is $n \times n$, computing its SVD is cheap.  The fourth step computes $QU$.  If $Q$ is not needed, i.e. only the singular vectors of $QU$ are desired, then we can pass $U$ to the third step and compute $QU$ directly without writing $Q$ to disk.  In this case, the SVD uses the same number of passes over the data as the QR factorization.  If only the singular values are needed, then only the first two steps of the algorithm are needed along with the SVD of $R$.  However, in this case, it would be favorable to use the TSQR implementation from Sec.~\ref{sec:indir_tsqr} to compute $R$.

One implementation challenge is matching the $Q$ and $R$ factors to the tasks on which they are computed.  In the first step, the key-value pairs emitted use a unique map task identifier (e.g., via the \texttt{uuid} package in Python) as the key and the $Q$ or $R$ factor as the value.  The reduce task in the second step maintains an ordered list of the keys read.  The $k$th key in the list corresponds to rows $(k - 1)n + 1$ to $kn$ of the locally computed $Q$ factor.  The map tasks in the third step parse a data file containing the $Q$ factors from the second step, and this redundant parsing allows us to skip the \emph{shuffle} and \emph{reduce}.  Another implementation challenge is that the map tasks in the first step and the reduce task in the second step must emit the $Q$ and $R$ factors to separate files.  For this functionality, we use the \texttt{feathers} extension of Dumbo.

\subsection{Extending Direct TSQR to a recursive algorithm}\label{sec:recursive}

A central limitation to the Direct TSQR method is the necessity of gathering all $R$ factors from the first step onto one processor in the second step.  As the matrix becomes fatter, this serial bottleneck becomes limiting.  We can cope with this issue by recursively extending the method with a recursive step following the first step.  The algorithm is outlined in Alg.~\ref{alg:MRDirectRecursive}.

\begin{algorithm}
  \caption{Recursive extension of direct method}
  \label{alg:MRDirectRecursive}
  \begin{algorithmic}
  \Function{DirectTSQR}{matrix A}
    \State Q1, R1 = FirstStep(A)
    \If{R1 is too big}
      \State Assign keys to rows of R1
      \State Q2 = DirectTSQR(R1)
    \Else
      \State Q2 = SecondStep(R1)
    \EndIf
    \State Q = ThirdStep(Q1, Q2)
    \State return Q
  \EndFunction
  \end{algorithmic}
\end{algorithm}

\section{Stability Experiments}\label{sec:stability}

\begin{figure}
\centering
\includegraphics[width=1\columnwidth]{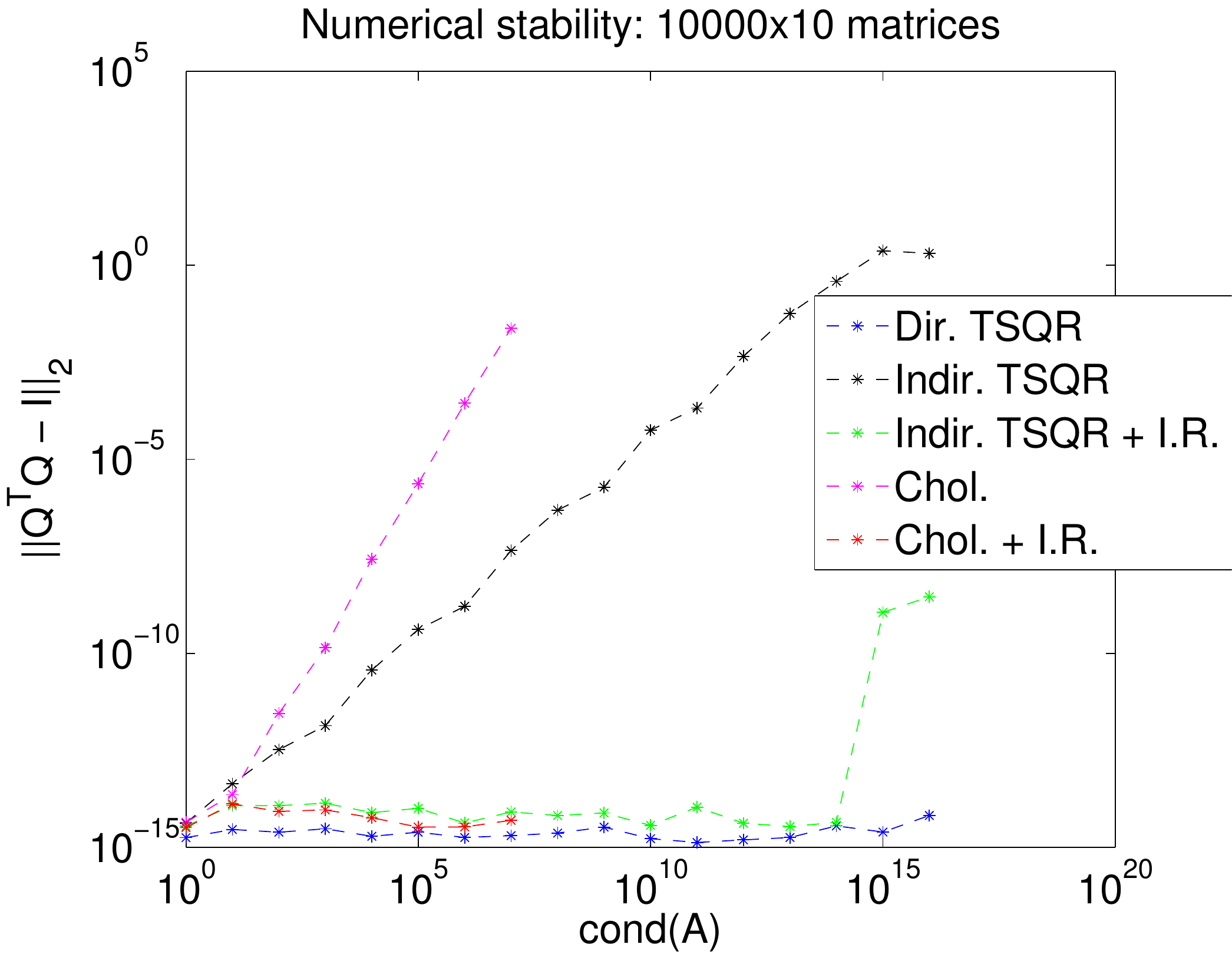}
\caption{Stability measurements for each algorithm for matrices of varying condition number}
\label{fig:stability}
\end{figure}

A major motivation for using the Direct TSQR method is numerical stability.  Based on prior work, we know that the Direct TSQR method should produce a matrix $Q$ with columns that are orthogonal to machine precision~\cite{demmel-2008-caqr2,Mori-2012-allreduce}, and Indirect TSQR and Cholesky QR should fail if the matrix is sufficiently ill-conditioned.  Fig.~\ref{fig:stability} shows results from a numerical stability experiment which measures the loss in orthogonality in $Q$ for Cholesky QR (with and without iterative refinement), Indirect TSQR (with and without iterative refinement),  and Direct TSQR.  We use $\normof[2]{Q^TQ - I}$ to measure the accuracy of $Q$.  As expected, using the inverse results in error that scales with the condition number.  One step of iterative refinement and the direct TSQR method both yield errors consistently around $10^{-15}$.  
Cholesky QR fails when the condition number of the matrix is $10^8$ or greater, and Indirect TSQR with iterative refinement has a large error when the condition number reaches $10^{16}$.  Previous work by Langou shows consistent results for similar experiments~\cite{langou2003-thesis}.

\section{Performance Experiments}\label{sec:perf}

We evaluate performance in three ways.  First, we build a performance model for our methods based on how much data is read and written by the MapReduce cluster.  Second, we evaluate the implementations on a 10-node, 40-core MapReduce cluster at Stanford's Institute for Computational and Mathematical Engineering (ICME).  Each node has 6 2-TB disks, 24 GB of RAM, and a single Intel Core i7-960 3.2 GHz processor.  They are connected via Gigabit ethernet.  After fitting only two parameters -- the read and write bandwidth -- the performance model predicts the actual runtime within a factor of two.  Finally, we explore the fault-tolerance of the MapReduce system by artificially introducing faults into each task.  Even when the frequency of faults is 1/8, the runtime only grows by about 23.2\%.

We do not perform standard parallel scaling studies due to how the Hadoop framework integrates the computational engine with the distributed filesystem. This combination makes these measurements difficult without rebuilding the cluster for each experiment.

\subsection{Performance model}\label{sec:perf_model}

Let $m_j$ and $r_j$ be the number of map and reduce tasks for step $j$, respectively.  Let $m_{max}$ be the maximum number of map tasks and $r_{max}$ be the maximum number of reduce tasks for the cluster.  Both $m_{max}$ and $r_{max}$ are fixed in the Hadoop configuration, and $m_{max}$ + $r_{max}$ is usually at least the total number of cores.  Let $k_j$ be the number of distinct input keys passed to the reduce tasks for step $j$.  We define the map parallelism for step $j$ as $p^m_j = \min\{m_{max}, m_j\}$ and the reduce parallelism for step $j$ as $p^r_j = \min\{r_{max}, r_j, k_j\}$.  Let $R^m_j$, $W^m_j$ be the amount of data read and written in the $j$th map task, respectively.  
We have analogous definitions for $R^r_j$ and $W^r_j$ for the $j$th reduce task.  Finally, let $\beta_r$ and $\beta_w$ be the inverse read and write bandwidth, respectively.  After computing $\beta_r$ and $\beta_w$, we can provide a lower bound for the algorithm by counting disk reads and writes.  The lower bound for a job with $N$ iterations is:
\[
T_{lb} = \ssum{j=1}{N}{\frac{R^m_j\beta_r + W^m_j\beta_w}{p^m_j} + \frac{R^r_j\beta_r + W^r_j\beta_w}{p^r_j}}.
\]

We use streaming benchmarks to estimate $\beta_r$ and $\beta_w$ for the 40-core ICME cluster, and the results are in Table~\ref{tab:streaming}.    On this cluster, $m_{max} = r_{max} = 40$.  Table~\ref{tab:reads_writes} provides the number of reads and writes for our algorithms, and Table~\ref{tab:parallelism_vars} provides the information for computing $p_j^m$ and $p_j^r$.  The keys for the matrix row identifiers are 32-byte strings.  The computed lower bounds for our algorithms are in Table~\ref{tab:lower_bounds}.  In Sec.~\ref{sec:perf_comparison}, we examine how close the implementations are to the lower bounds. 

\begin{table*}[tbp]
\vspace{-\baselineskip}
\centering
\caption{Streaming time to read from and write to disk.  Performance is in inverse bandwidth, so larger $\beta_r$ and $\beta_w$ means slower streaming.  The streaming benchmarks are performed with $m_{max}$ map tasks.}
\begin{tabular}{l l c @{\qquad\qquad} c c @{\qquad\qquad} c c}
\toprule
Rows & Cols. & HDFS Size & read+write & read & $\beta_r/m_{max}$ & $\beta_w/m_{max}$ \\
           &            &    (GB)         &     (secs.)     & (secs.) & (s/GB)    & (s/GB) \\
\midrule
4,000,000,000 & 4     & 134.6 & 713 & 305 & 2.266 & 3.0312 \\
2,500,000,000 & 10   & 193.1 & 909 & 309 & 1.6002 & 3.1072 \\
600,000,000    & 25   & 112.0 & 526 & 169 & 1.5089 & 3.1875 \\
500,000,000    & 50   & 183.6 & 848 & 253 & 1.378 & 3.2407 \\
150,000,000    & 100 & 109.6 & 504 & 152 & 1.3869 & 3.2117 \\
\bottomrule
\end{tabular}
\label{tab:streaming}
\end{table*}

\begin{table*}[tbp]
\setlength{\extrarowheight}{1.5pt}
\centering
\caption{Number of reads and writes at each step (in bytes).  We assume a double is 8 bytes and $K$ is the number of bytes for a row key ($K = 32$ in our experiments).  Only one iteration of Householder QR is shown: the lower bound repeats this iteration $n$ times.  The amount of key data is separated from the amount of value data.  For example, $8mn + Km$ is $Km$ bytes in key data and $8mn$ bytes in value data.}
\begin{tabular}{c @{\qquad} c c c c}
\toprule
 & Cholesky & Indirect & Direct & House. \\
 &                   & TSQR    & TSQR & (1 step) \\            
 \midrule
$R_1^m$ & $8mn + Km$ & $8mn + Km$ & $8mn + Km$ & $8mn +Km$ \\
$W_1^m$ & $8m_1n^2 + 8m_1n$ & $8m_1n^2 + 8m_1n$ & $8mn + 8m_1n^2 + Km + 64m_1$ & $8mn + Km$ \\

$R_1^r$ & $8m_1n^2 + 8m_1n$ & $8m_1n^2 + 8m_1n$ & $0$ & $0$\\
$W_1^r$ & $8n^2 + 8n$ & $8r_1n^2 + 8r_1n$ & $0$ & $0$\\

$R_2^m$ & $8n^2 + 8n$ & $8r_1n^2 + 8r_1n$ &  $8m_1n^2 + Km_1$ & $8mn + Km$ \\
$W_2^m$ & $8n^2 + 8n$ & $8r_1n^2 + 8r_1n$ & $8m_1n^2 + Km_1$ & $16m_1$\\

$R_2^r$ & $8n^2 + 8n$ & $8r_1n^2 + 8r_1n$ & $8m_1n^2 + Km_1$ & $0$ \\
$W_2^r$ & $8n^2 + 8n$ & $8n^2 + 8n$ & $8m_1n^2 + 32m_1 + 8n^2 + 8n$ & $0$ \\

$R_3^m$ & $8mn + Km + m_3(8n^2 + 8n)$ & $8mn + Km + m_3(8n^2 + 8n)$ & $8mn + Km + m_3(8m_1n^2 + 64m_1)$ & ---\\
$W_3^m$ & $8mn + Km$ & $8mn + Km$ & $8mn + Km$ & ---\\

$R_3^r$ & $0$ & $0$  & $0$ & --- \\ 
$W_3^r$ & $0$ & $0$ & $0$ & ---\\
\bottomrule
\end{tabular}
\label{tab:reads_writes}
\end{table*}

\begin{table}[tbp]
\setlength{\extrarowheight}{0.5pt}
\centering
\caption{Values needed to compute $p_j^m$ and $p_j^r$.  For Householder QR, only the data for one step is shown.  Each step of Householder QR has identical data.  Both   $m_1$ and $m_3$ are dependent on the matrix size.  Other listed data are not.}
\begin{tabular}{l l c c c c}
\toprule
\multicolumn{2}{l}{} & Cholesky & Indirect & Direct & House. \\
& & & TSQR & TSQR & (1 step) \\
 \midrule
             $4.0B \times 4$ & $m_1$ & $1200$ & $1200$ & $2000$ & $1200$ \\
              $2.5B \times 10$ & & $1680$ & $1680$ & $2640$ & $1680$ \\
              $600M \times 25$ & & $1200$ & $1200$ & $1600$ & $1920$ \\
              $500M \times 50$ & & $1920$ & $1920$ & $2560$ & $1920$ \\
              $150M \times 100$ & & $1200$ & $1200$ & $1600$ & $1200$ \\ \midrule
& $m_2$ & $m_{max}$ & $m_{max}$ & $m_{max}$ & --- \\ \midrule
              $4.0B \times 4$ & $m_3$ & $1200$ & $1200$ & $2000$ & --- \\
              $2.5B \times 10$ & & $1680$ & $1680$ & $2640$ & --- \\
              $600M \times 25$ & & $1200$ & $1200$ & $1600$ & --- \\
              $500M \times 50$ & & $1920$ & $1920$ & $2560$ & --- \\
              $150M \times 100$ & & $1200$ & $1200$ & $1600$ & --- \\
\midrule
& $r_1$ & $r_{max}$ & $r_{max}$ & $r_{max}$ & --- \\
& $r_2$ & $1$ & $1$ & $1$ & --- \\ 
 \midrule
& $k_1$ & $n$ & $m_1n$ & $m_1$ & ---\\
& $k_2$ & $n$ & $m_1n$ & $m_1$ & ---\\
& $k_3$ & $0$ & $0$ & $0$ & ---\\
\bottomrule
\end{tabular}
\label{tab:parallelism_vars}
\end{table}

\begin{table*}[tbp]
\vspace{-\baselineskip}
\caption{Computed lower bounds for each algorithm.}
\centering
\begin{tabularx}{\linewidth}{llXXXXXXXX}
\toprule
Rows & Cols. & Cholesky & Indirect & Cholesky & Indirect         & Direct & House. \\
           &           &                   & TSQR   & +I.R.         & TSQR+I.R.    & TSQR & \\\cmidrule{3-8}
& & \multicolumn{2}{l}{$T_{lb}$ (secs.)} \\ \midrule
4,000,000,000 & 4   & 1803  & 1803 & 3606 & 3606 & 2528 & 7213\\
2,500,000,000 & 10 & 1645  & 1645 & 3290 & 3290 & 2464 & 16448\\
600,000,000 & 25    & 804    & 804    & 1609 & 1609 & 1236 & 20111\\
500,000,000 & 50    & 1240  & 1240 & 2480 & 2480 & 2095 & 61989\\
150,000,000 & 100  & 696    & 696   & 1392 & 1392 & 1335 & 69569\\
\bottomrule
\end{tabularx}
\label{tab:lower_bounds}
\end{table*}

\subsection{Algorithmic comparison}\label{sec:perf_comparison}

Using one step of iterative refinement yields numerical errors that are acceptable in a vast majority of cases.  In these cases, performance is our motivator for algorithm choice.  Tabs. \ref{tab:raw_performance} and \ref{tab:work_performance} show performance results of the Indirect and Direct TSQR methods, Cholesky QR, and Householder QR for a variety of matrices.  The running time of Householder QR is long enough that we extrapolate the performance data from the first four steps of the algorithm.

\begin{table*}[tbp]
\vspace{-\baselineskip}
\caption{Times to compute $QR$ on a variety of matrices with four MapReduce algorithms.  *Householder QR data extrapolated from the first four steps of the algorithm.}
\centering
\begin{tabularx}{\linewidth}{llXXXXXXX}
\toprule
Rows & Cols. & HDFS Size & Cholesky & Indirect & Cholesky & Indirect       & Direct & House.*\\
           &           & (GB)             &                   & TSQR    & +I.R.         & TSQR+I.R. & TSQR & \\ \cmidrule{4-9}
& & & \multicolumn{2}{l}{job time (secs.)} \\ \midrule
4,000,000,000 & 4     & 134.6 & 2931 & 4076  & 5832  & 7431 & 6128 & 15021\\
2,500,000,000 & 10   & 193.1 & 2508 & 2509  & 5011  & 5052 & 4035 & 32950 \\
600,000,000    & 25   & 112.0 & 1098 & 1104  & 2221  & 2235 & 1910 & 37388 \\
500,000,000    & 50   & 183.6 & 1563 & 1618  & 3204  & 3298 & 3090 & 117775\\
150,000,000    & 100 & 109.6 & 921   & 954    & 1878  & 1960 & 2154 & 133025\\
\bottomrule
\end{tabularx}
\label{tab:raw_performance}
\end{table*}

\begin{table*}[tbp]
\vspace{-\baselineskip}
\centering
\caption{Floating point operations per second on a variety of matrices with four MapReduce algorithms.}
\begin{tabularx}{\linewidth}{llXXXXXXX}
\toprule
Rows & Cols. & $2*$rows$*$cols$^2$ & Cholesky & Indirect & Cholesky & Indirect       & Direct & House.*\\
           &           &                                                  &                   & TSQR    & +I.R.         & TSQR+I.R. & TSQR & \\ \cmidrule{4-9}
& & & \multicolumn{2}{l}{$2*$rows$*$cols$^2/$sec} \\ \midrule
4,000,000,000 & 4     & 1.28e+11 &  4.37e+07 &   3.14e+07 & 2.19e+07 &  1.72e+07  & 2.09e+07 &  8.52e+06\\
2,500,000,000 & 10   & 5.00e+11 &  1.99e+08 &  1.99e+08  & 9.98e+07  & 9.90e+07   & 1.24e+08 & 1.52e+07\\
600,000,000    & 25   & 7.50e+11 &  6.83e+08 &  6.79e+08  & 3.38e+08  & 3.36e+08   & 3.93e+08 & 2.01e+07\\
500,000,000    & 50   & 2.50e+12 &  1.60e+09 &  1.55e+09  & 7.80e+08  & 7.58e+08   & 8.09e+08 & 2.12e+07\\
150,000,000    & 100 & 3.00e+12 &  3.26e+09 &  3.14e+09  & 1.60e+09  & 1.53e+09  & 1.39e+09 & 2.26e+07\\
\bottomrule
\end{tabularx}
\label{tab:work_performance}
\end{table*}

In our experiments, we see that Indirect TSQR and Cholesky QR provide the fastest ways of computing the $Q$ and $R$ factors, albeit $\normof[2]{Q^TQ - I}$ may be large.  For all matrices with greater than four columns, these two methods have similar running times.  For such matrices, the majority of the running time is the $AR^{-1}$ step, and this step is identical between the two methods.  This is precisely because the write bandwidth is less than the read bandwidth.

For the matrices with 10, 25, and 50 columns, Direct TSQR outperforms the indirect methods with iterative refinement.  The performance gain for this method is the greatest for smaller numbers of columns.  However, when the matrix becomes too skinny (e.g., with four columns), Cholesky QR with iterative refinement is a better choice.  When the matrix becomes too fat (e.g., with 100 columns), the local gather in Step 2 becomes expensive.  Table~\ref{tab:perf_parts} shows the amount of time spent in each step of the Direct TSQR computation.  Indeed, Step 2 consumes a larger fraction of the running time as the number of columns increases.

For every matrix, Householder QR is by far the slowest method.  As the number of columns grows, the algorithm becomes continuously less competitive.

\begin{table}[tbp]
\vspace{-\baselineskip}
\caption{Fraction of time spent in each step of the Direct TSQR algorithm (fractions may not sum to 1 due to rounding).}
\centering
\begin{tabular}{l l @{\qquad\qquad} c c c}
\toprule
Rows & Cols. & Step 1 & Step 2 & Step 3 \\
\midrule
4,000,000,000 & 4     &  0.72 & 0.02 & 0.26 \\
2,500,000,000 & 10   & 0.61 & 0.04  & 0.34 \\
600,000,000    & 25   & 0.56 & 0.06 & 0.38 \\
500,000,000    & 50   & 0.55 & 0.07 & 0.39 \\
150,000,000    & 100 & 0.47 & 0.15 & 0.38 \\
\bottomrule
\end{tabular}
\label{tab:perf_parts}
\end{table}

Table~\ref{tab:bound_perf} shows how each algorithm performs compared to its lower bound from Table~\ref{tab:lower_bounds}.  We see that Direct TSQR diverges from this bound when the number of columns is too small.  To explain this difference, we note that Direct TSQR must gather all the keys and values in the first step before performing any computation.  When the number of key-value pairs is large, e.g., the $4,000,000,000 \times 4$ matrix, then this step becomes limiting and this is not accounted for by our performance model.
Thus, the model predicts the runtime of Cholesky QR and Indirect TSQR with iterative refinement more accurately than Direct TSQR.  Although their lower bounds are greater, the empirical performance makes these algorithms more attractive as the number of columns increases.  The enormous lower bound of Householder QR makes the algorithm entirely unattractive, which renders Direct TSQR the best algorithm if guaranteed stability is required.

\begin{table*}[tbp]
\vspace{-\baselineskip}
\caption{Performance of algorithms as a multiple of the lower bounds from Table~\ref{tab:lower_bounds}.}
\centering
\begin{tabularx}{\linewidth}{llXXXXXXXXX}
\toprule
Rows               & Cols. & Cholesky & Indirect & Cholesky & Indirect         & Direct & House. \\
                          &           &                   & TSQR   & +I.R.         & TSQR+I.R. & TSQR & \\\cmidrule{3-8}
& & multiple of $T_{lb}$ \\ \midrule
4,000,000,000 & 4 & 1.6256 & 2.2607 & 1.6173 & 2.0607 & 2.4241 & 2.0825 \\
2,500,000,000 & 10 & 1.5246 & 1.5252 & 1.5231 & 1.5356 & 1.6376 & 2.0033 \\
600,000,000    & 25 & 1.3657 & 1.3731 & 1.3804 & 1.3891 & 1.5453 & 1.8591 \\
500,000,000    & 50 & 1.2605 & 1.3048 & 1.2919 & 1.3298 & 1.4749 & 1.8999 \\
150,000,000   & 100 & 1.3233 & 1.3707 & 1.3491 & 1.4080  & 1.6135 & 1.9121 \\
\bottomrule
\end{tabularx}
\label{tab:bound_perf}
\end{table*}

\subsection{Fault tolerance}

\begin{figure}
\includegraphics[width=1.0\columnwidth]{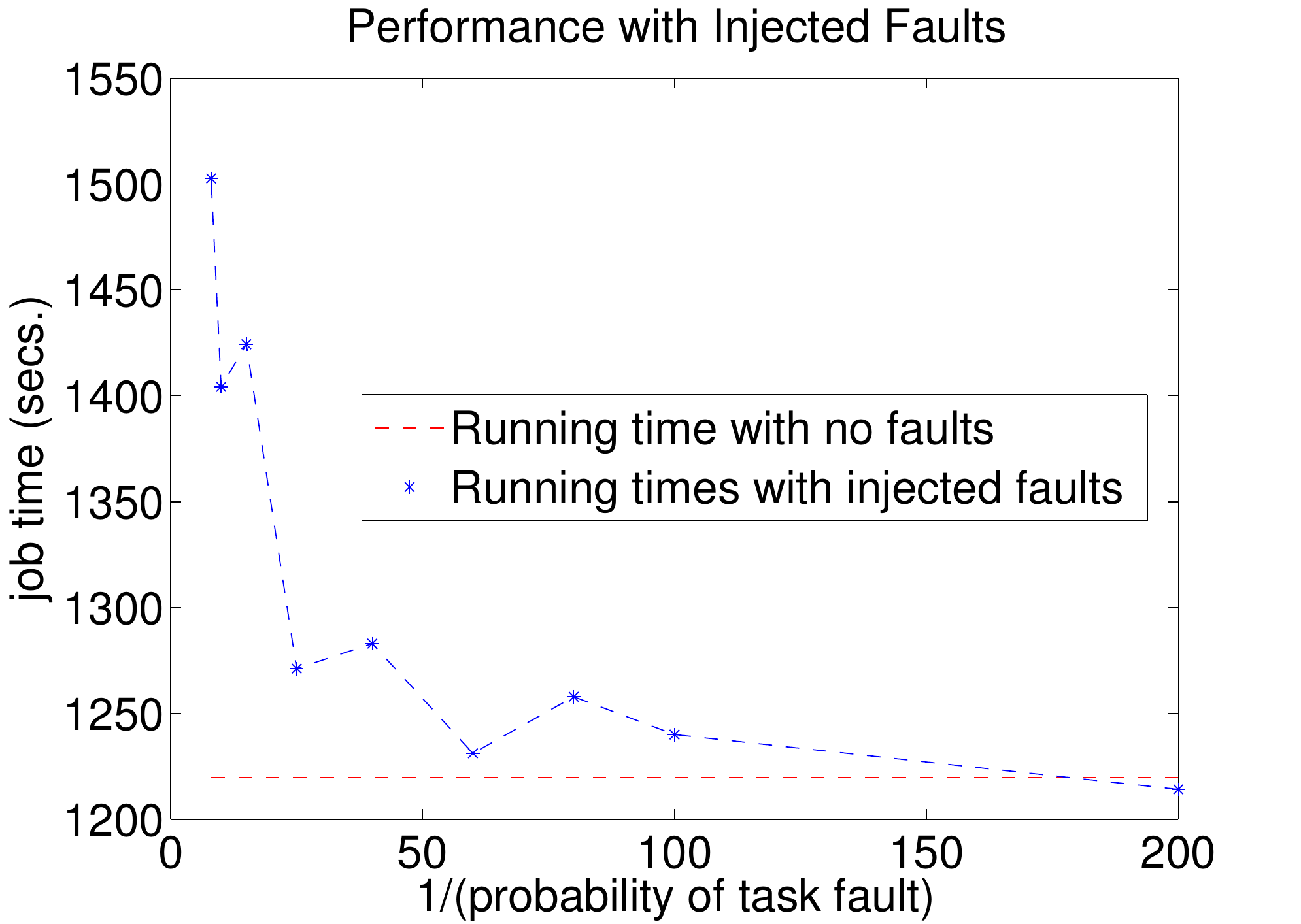}
\caption{Running time of Direct TSQR on an $800,000,000 \times 10$ matrix with injected faults}
\label{fig:faults}
\end{figure}

One motivation for using a MapReduce architecture is fault tolerance.  We measure the effects of faults on performance by crashing tasks with a certain probability of fault.  Fig.~\ref{fig:faults} shows how the performance changes as we vary the probability of failure for tasks while running the Direct TSQR method on a matrix with 800 million rows and 10 columns.  This matrix occupies 62.9 GB on HDFS.

In total, 800 map tasks are launched for each map stage of the Direct TSQR method.  With no injected faults, the running time is 1220 seconds.  When the probability of a fault is 1/8, the running time is 1503 seconds, only a 23.2 \% performance penalty.

\section{Conclusion}

If numerical stability is required, the Direct TSQR method discussed in this paper is the best choice of algorithm.  It is guaranteed to produce a numerically orthogonal matrix.  It usually takes no more than twice the time of the fastest, but unstable method, and it often outperforms conceptually simplier methods. It is also orders of magnitude faster than the Householder QR method implemented in MapReduce. 

All of the code used for this paper is openly available online, see: \begin{center}
\url{https://github.com/arbenson/mrtsqr} \end{center}
This software runs on any system supporting Hadoop streaming, including cluster management systems like Mesos~\cite{Mesos2011}.  

In the future we plan to investigate mixed MPI and Hadoop code.  The idea is that once all the local mappers have run in the first step of the Direct TSQR method, the resulting $R_i$ matrices constitute a much smaller input.  If we run a standard, in-memory MPI implementation to compute the QR factorization of this smaller matrix, then we could remove two iterations from the direct TSQR method.  Also, we would remove much of the disk IO associated with saving the $Q_i$ matrices.  We believe these changes would make our MapReduce codes significantly faster.

\section*{Acknowledgment}

Austin Benson is supported by an Office of Technology Licensing Stanford Graduate Fellowship.  Many implementation optimizations were done by Austin Benson for the CS 267 (instructed by James Demmel and Kathy Yelick) and Math 221 (instructed by James Demmel) courses at UC-Berkeley.  Thanks to the team at NERSC, including Lavanya Ramakrishnan and Shane Canon, for help with MapReduce codes on the Magellan cluster.

David F.~Gleich is supported by a DOE CSAR grant.

Research supported by Microsoft (Award \#024263) and Intel (Award \#024894) funding and by matching funding by U.C. Discovery (Award \#DIG07-10227). Additional support comes from Par Lab affiliates National Instruments, Nokia, NVIDIA, Oracle, and Samsung.  Research is also supported by DOE grants DESC0003959 and DE-SC0004938.

We are grateful to Stanford's Institute for Computational and Mathematical Engineering for letting us use their MapReduce cluster for these computations.

We are grateful to Paul Constantine for working on the initial TSQR method and for continual discussions about using these routines in simulation data analysis problems.

\bibliographystyle{abbrv}
\bibliography{all-bibliography}

\end{document}